# CXCR6, a Newly Defined Biomarker of Tissue-Specific Stem Cell Asymmetric Self-Renewal, Identifies More Aggressive Human Melanoma Cancer Stem Cells

Rouzbeh Taghizadeh[1], Minsoo Noh[2], Yang Hoon Huh[1], Emilio Ciusani[3], Luca Sigalotti[4], Michele Maio[4], Beatrice Arosio[5], Maria R. Nicotra[6], PierGiorgio Natali[7], James L. Sherley[1,9], Caterina A. M. La Porta[8*,9]

1 Programs in Regenerative Biology and Cancer Biology, Adult Stem Cell Technology Center, Boston Biomedical Research Institute, Watertown, Massachusetts, United States of America, 2 College of Pharmacy, Ajou University, Suwon, Republic of Korea, 3 Istituto Neurologico Carlo Besta, Milano, Italy, 4 Cancer Bioimmunotherapy Unit, Department of Medical Oncology, Centro di Riferimento Oncologico, Istituto di Ricovero e Cura a Carattere Scientifico, Aviano, Italy, 5 Department of Internal Medicine, Università degli Studi di Milano, Milano, Italy, 6 Molecular Biology and Pathology Institute, National Research Council, Roma, Italy, 7 CINBO Laboratory, University Chieti, Chieti, Italy, 8 Department of Biomolecular Science and Biotechnology, University of Milan, Milan, Italy

## Abstract

***Background:*** A fundamental problem in cancer research is identifying the cell type that is capable of sustaining neoplastic growth and its origin from normal tissue cells. Recent investigations of a variety of tumor types have shown that phenotypically identifiable and isolable subfractions of cells possess the tumor-forming ability. In the present paper, using two lineage-related human melanoma cell lines, primary melanoma line IGR39 and its metastatic derivative line IGR37, two main observations are reported. The first one is the first phenotypic evidence to support the origin of melanoma cancer stem cells (CSCs) from mutated tissue-specific stem cells; and the second one is the identification of a more aggressive subpopulation of CSCs in melanoma that are CXCR6+.

***Methods/Findings:*** We defined CXCR6 as a new biomarker for tissue-specific stem cell asymmetric self-renewal. Thus, the relationship between melanoma formation and ABCG2 and CXCR6 expression was investigated. Consistent with their non-metastatic character, unsorted IGR39 cells formed significantly smaller tumors than unsorted IGR37 cells. In addition, ABCG2+ cells produced tumors that had a 2-fold greater mass than tumors produced by unsorted cells or ABCG2- cells. CXCR6+ cells produced more aggressive tumors. CXCR6 identifies a more discrete subpopulation of cultured human melanoma cells with a more aggressive MCSC phenotype than cells selected on the basis of the ABCG2+ phenotype alone.

***Conclusions/Significance:*** The association of a more aggressive tumor phenotype with asymmetric self-renewal phenotype reveals a previously unrecognized aspect of tumor cell physiology. Namely, the retention of some tissue-specific stem cell attributes, like the ability to asymmetrically self-renew, impacts the natural history of human tumor development. Knowledge of this new aspect of tumor development and progression may provide new targets for cancer prevention and treatment.

**Funding:** This work was supported in part by the National Institues of Health Director's Pioneer Award #5DP1OD000805 and by National Italian grant COFIN 2008. The funders had no role in study design, data collection and analysis, decision to publish, or preparation of the manuscript.

**Competing Interests:** The authors have declared that no competing interests exist.

* E-mail: caterina.laporta@unimi.it

9 These authors contributed equally to this work.

## Introduction

Cancer chemotherapy efficacy is frequently impaired by either intrinsic or acquired tumor resistance, a phenomenon termed multi-drug resistance (MDR) [1]. MDR can result from several distinct mechanisms, including reducing drug accumulation in tumor cells [1]. The mechanism that is most commonly encountered in the laboratory is the increased efflux of a broad class of hydrophobic cytotoxic drugs that is mediated by one of a family of energy-dependent transporters (ABC family) [2]. Although several members of the superfamily have dedicated particular functions involving the transport of specific substrates, it is becoming increasingly evident that the complex physiological network of ABC transporters has a pivotal role in host detoxification and protection of the body against xenobiotics. Among the human ABC superfamily, only ABCB1, ABCC1 (MDR1) and ABCG2 have to date been shown to mediate MDR, each with distinct overlapping efflux substrate specificities and tissue distribution patterns [3]. Fulfilling their role in detoxification, several ABC transporters have been found to be overexpressed in cancer cell lines cultured under selective pressure. In particular, ABCB5 is overexpressed in melanoma, and ABCG2 is expressed by a subcellular CD133-positive melanoma cells [4,5].

In this report, we investigated a new candidate for a melanoma cancer stem cell (MCSC) marker, the cytokine co-receptor CXCR6, with respect to the properties of ABCG2-expressing

MCSCs. An important unresolved question in the field of cancer stem cell (CSC) research is whether there is a direct lineage relationship between CSCs and tissue-specific stem cells (TSSCs) found in the normal tissues from which cancers arise. It is worthwhile to identify markers that distinguish between tumorigenic from non tumorigenic cells [6]. CD133 is expressed by most of melanoma cell lines [4], and it does not seem able to distinguish tumorigenic from non -tumorigenic cells [6]. Moreover recently, Weissman's group confirmed the presence of a more aggressive subpopulation CD217+ in human melanoma [8]. We set out to shed experimental light on this issue by investigating previously described melanoma CSCs associated with established melanoma cell lines [4,6] for evidence of asymmetric self-renewal, a specific property of TSSCs [7,9]. For this evaluation, we employed a new biomarker that we show here to be associated with asymmetric self-renewal, the chemokine receptor CXCR6. In earlier studies, CXCR6 was identified as a co-receptor for human immunodeficiency virus infection of lymphocytes [10]. Low levels of CXCR6 expression have also been detected specifically on memory/effector Th1 cells in peripheral blood [11]. More recently, expression of CXCR6 was associated with human tumors, including melanoma [12–14]. Herein, we show, based on studies with mouse cell lines genetically engineered to undergo asymmetric self-renewal like TSSCs, that both the pattern and level of CXCR6 expression is specifically associated with asymmetric self-renewal division.

We looked for co-association of CXCR6 expression and ABCG2 expression with MCSC activity and evidence for the descent of melanoma CSCs from TSSCs. In fact, we find that a large percentage of ABCG2-positive melanoma cells with known CSC properties, also express CXCR6. Moreover, like ABCG2-expressing melanoma cells, CXCR6-expressing cells are a small fraction of cells in cultures of primary, metastatic melanoma cell lines and human melanoma biopsy specimens. *In vivo* CXCR6+ melanoma cells produced a tumor in less time than ABCG2+ cells and, more interestingly, the negative subpopulation did not give a tumor. This phenotypic co-association of CXCR6, a biomarker for asymmetric self-renewal in non-cancerous cells, with CSC activity in tumor-derived cells is evidence for a direct lineage relationship between TSSCs and CSCs in the case of melanocytic cells and melanoma, respectively.

Finally, we have characterized the subpopulation ABCG2+/ABCG2- and CXCR6+/CXCR6- of melanoma for the expression of melanoma associated antigens (MAA) in view of a possible immunotherapeutic approach against more aggressive subpopulations expressing ABCG2 and/or CXCR6.

## Results

### Identification of CXCR6 as a biomarker for asymmetric self-renewal division

Previously, we have reported studies with a genetically engineered cultured cell model that provides experimentally controlled asymmetric self-renewal divisions similar to those of TSSCs [15,16,17]. The cells contain a Zn-regulated wild-type p53 cDNA gene. In Zn-free medium, these cells divide with symmetric cell kinetics, producing two cycling cells from each division. These divisions model symmetric TSSC divisions in which two stem cells are produced. In $ZnCl_2$-supplemented culture medium, consequent normal levels of p53 expression induce asymmetric self-renewal divisions that are characterized by divisions that produce one cycling sister and one sister that arrests at G1/S. These divisions model asymmetric TSSC divisions that produce one stem cell and one cell that either differentiates or serves as the progenitor for a differentiating cell lineage.

In gene micro-array analyses (NCBI-GEO data base, http://www.ncbi.nlm.nih.gov/geo/query/acc.cgi?token = dlmlzmuowumsqxy& acc = GSE25334), *Cxcr6* was identified as a gene whose expression was undetectable in symmetrically self-renewing congenic p53-null control cells grown in $ZnCl_2$-supplemented medium. However, it was induced in p53-inducible cells undergoing asymmetric self-renewal under the same culture conditions. Quantitative RT-PCR analyses revealed that symmetrically self-renewing cells did express *Cxcr6* mRNA at a detectable level; but its expression increased $8.6 \pm 4.4$ fold (n = 6 analyses; p = 0.002) in cells induced to shift to asymmetric self-renewal.

The initial micro-array analyses also showed that asymmetric self-renewal was obligatory for the increase in *Cxcr6* expression; and thereby indicated that the increased expression was not due to p53 expression alone. *Cxcr6* expression was also undetectable in congenic p53-inducible cells genetically engineered with forced expression of the type II inosine monophosphate dehydrogenase gene (IMPDH II), even when p53 expression was induced. Expression of IMPDH II, the rate-limiting enzyme for cellular guanine ribonucleotide biosynthesis, prevents p53-induced asymmetric self-renewal, while leaving p53-dependent gene regulation intact [16,18,19]. By this criterion, *Cxcr6* was defined as a specific "asymmetric self-renewal associated (ASRA)" gene.

To investigate the ASRA biomarker properties of CXCR6 protein, we examined its pattern of expression determined by indirect *in situ* immunofluorescence (ISIF) detection in two related single-cell self-renewal pattern assays. In the first, the sister pair assay (SP; [19]), cells are plated at a sufficiently low density to allow delineation of sister cells by their relatively closer proximity after division. In the second, cells are analyzed after division in the presence of cytochalasin D (CD; [20]). CD prevents cytokinesis, but does not prevent nuclear division. The binucleated cells thereby produced provide a more confident comparison of sister nuclei.

In previous applications of the SP (previously called "daughter pair;" [19] and CD assays, bromodeoxyuridine (BrdU) incorporation was used as a marker for cycling S phase cells. For the present studies, we used the cell cycle phase-specific protein cyclin A as an indicator of cycling cells. Cyclin A is a well-described marker of cycling cells that increases progressively in level from late G1 phase to G2 phase of the cell cycle [21]. The epifluorescence micrographs in Fig. 1 show how indirect ISIF can be employed in SP and CD assays to distinguish asymmetrically self-renewing cells (ASYM) from symmetrically renewing ones (SYM). Under conditions that foster symmetric self-renewal, sister cell pairs and CD-arrested binucleated cells show a high frequency of symmetric cyclin A detection. In contrast, under conditions that foster asymmetric self-renewal, an asymmetric pattern of cyclin A expression is significantly more frequent.

As shown in Fig. 1, CXCR6 protein is detected in both the cytoplasm and nucleus of the genetically engineered cells used to identify CXCR6 as an ASRA biomarker. The level of nuclear expression is higher in cells under conditions of asymmetric self-renewal, consistent with the mRNA expression data, which was described earlier. In both SP and CD analyses, under conditions of symmetric self-renewal, CXCR6 protein is detected in both nuclei of sister cells. In contrast, under conditions that promote asymmetric self-renewal, cells show a higher frequency of the asymmetric expression pattern that is also directly correlated with the asymmetric pattern of cyclin A. In quantitative CD analyses, under conditions for symmetric self-renewal, only 6% of binucleated cells showed coordinate asymmetric expression of cyclin A and CXCR6; whereas 27% showed this specific pattern under conditions for asymmetric self-renewal (p = 0.001). These findings identify CXCR6 as a first-described up-regulated biomarker of cells undergoing asymmetric self-renewal like TSSCs.

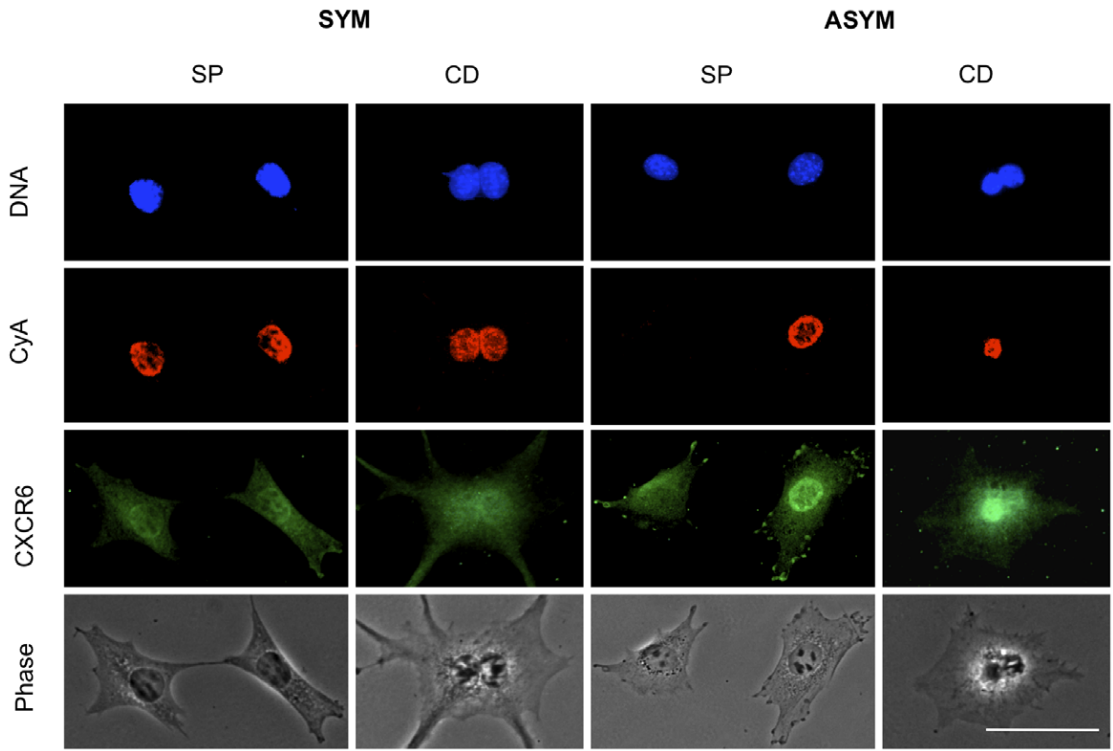

**Figure 1. Demonstration of the asymmetric self-renewal associated (ASRA) biomarker properties of CXCR6.** Shown are examples of epifluorescence and phase contrast micrographs from parallel **SP** and **CD** analyses, as described in the text. **SYM**, symmetric self-renewal (Congenic p53-null engineered cell lines grown in $ZnCl_2$-supplemented medium). **ASYM**, asymmetric self-renewal (Zn-dependent, p53-inducible engineered cells grown in $ZnCl_2$-supplemented medium). **DNA**, DAPI nuclear DNA fluorescence. **CyA**, indirect ISIF with specific antibodies for cyclin A. **CXCR6**, indirect ISIF with specific antibodies for CXCR6. Note that in the ASYM state, CXCR6 is up-regulated in the cycling cell, which models asymmetrically self-renewing TSSCs. **Phase**, phase contrast micrograph. **Scale bar** = 50 microns.
doi:10.1371/journal.pone.0015183.g001

### ABCG2 and melanoma formation

In a recent paper, our group published that ABCG2 is expressed by a subpopulation of human melanoma cells (cell line WM115) expressing high level of CD133 [4]. We extended this analysis of ABCG2 to two additional melanoma cell lines derived from the same melanoma patient: primary melanoma cell line IGR39 and the related metastatic line IGR37.

First, as shown in Table 1, we analyzed three polymorphisms C/A (421 C>A), C/T (S248T) and C/T (F208S) in the unsorted populations and in the ABCG2+ and ABCG2- sorted subpopulations revealing a heterozygous status of ABCG2 for all the three polymorphisms. These results are consistent with the origin of the cell lines from the same patient. We focused on these polymorphisms since SNP 421C>A is most common between different ethnic groups [22] and has been associated with decreased expression of the ABCG2 protein [23]. Moreover, F208S and S248P polymorphisms were associated with defective active transport of methotrexate and haematoporphyrin [24]; and, in particular, F208S variant proteins (both-glycosylated and immature forms) are recognized as misfolded proteins by putative "check point" systems and readily undergo ubiquination and protein degradation in proteasomes [25].

Respectively, ~11% and ~40% of cells in actively growing cultures of IGR37 and IGR39 had detectable expression for ABCG2 (Fig. 2). IGR37 and IGR39 cells were sorted into ABCG2+ and ABCG2- subpopulations and transplanted into NOD-SCID mice (n = 3 for each subpopulation evaluated). Unsorted cells were also injected into NOD-SCID mice (n = 3) for comparison. After two months, mice from all cohorts were sacrificed. Tumor masses were excised, imaged, weighed, and digested to isolate single cell suspensions for further analysis. As shown in Table 2 and Fig. 3, the transplantation of ABCG2+ IGR37 cells resulted in tumors with 2-fold greater mass on average compared to tumors that arose from ABCG2- cells (p<0.01). Interestingly, injection of unsorted IGR37 cells also result in smaller tumors than ABCG2+ cells that were statistically similar in sizes to tumors from ABCG2- cells (Fig. 3; Table 2). Unsorted IGR39 cells resulted in smaller tumor masses, compared to

**Table 1.** Analysis of ABCG2 polymorphisms.

| Cells | 421 C>A (C/A) | S248P (C/T) | F208S (C/T) |
| --- | --- | --- | --- |
| IGR37 | CA | CT | CT |
| IGR39 | CA | CT | CT |
| ABCG2+ IGR37 | CA | CT | CT |
| ABCG2- IGR37 | CA | CT | CT |
| ABCG2+ IGR39 | CA | CT | CT |
| ABCG2- IGR39 | CA | CT | CT |

doi:10.1371/journal.pone.0015183.t001

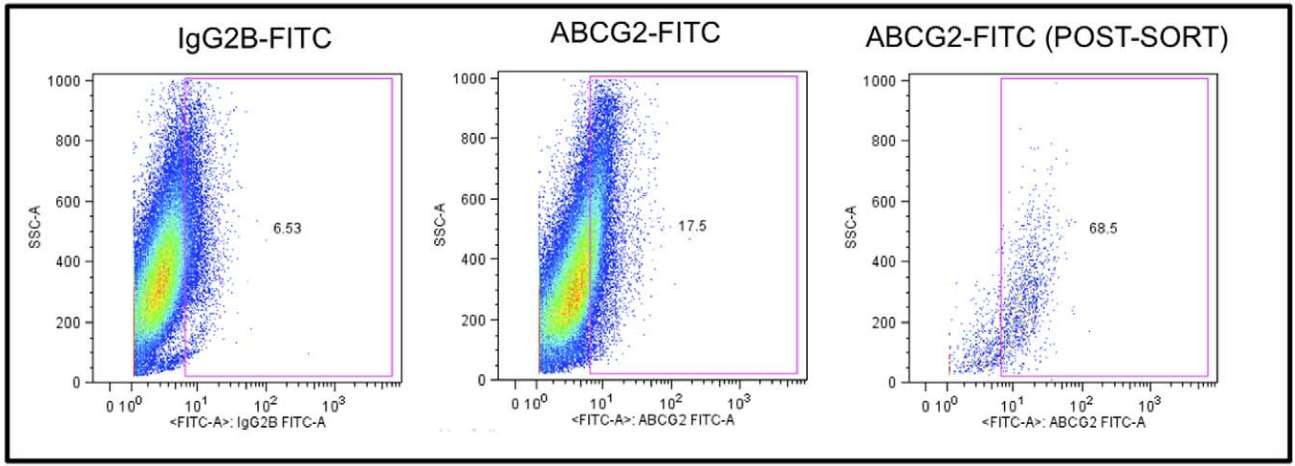

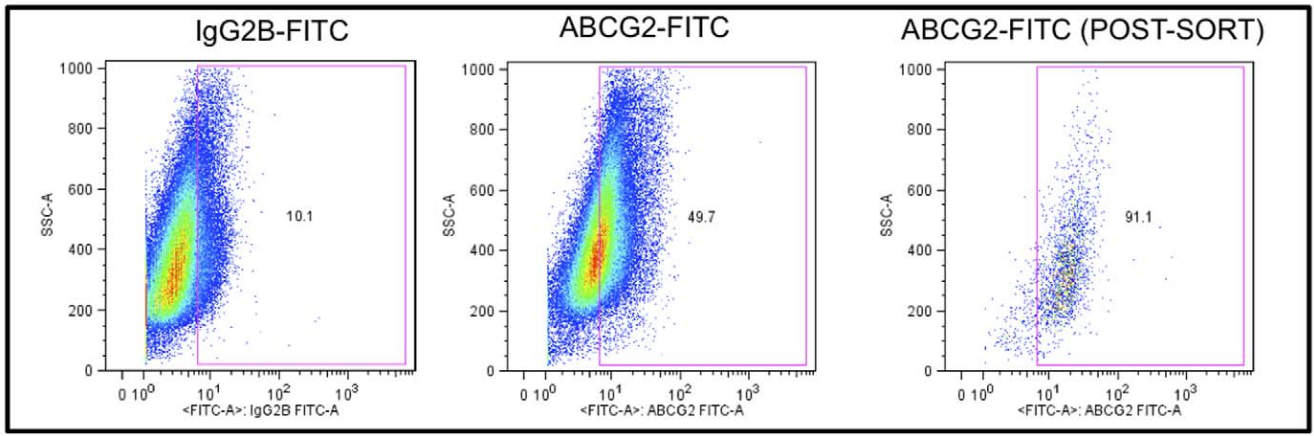

**Figure 2. Flow cytometry detection and sorting of ABCG2+ subpopulations from cultures of human melanoma cell lines.** Primary melanoma IGR39 cells and metastatic melanoma IGR37 cells were incubated with the indicated fluorescent FITC-conjugated antibodies and analyzed by flow cytometry as described in *Materials and Methods*. **Y-axes**, side scatter; **x-axes**, relative fluorescence intensity. **Left panels**, analyses with isotype (IgG) control antibodies; **middle panels**, analyses with anti-ABCG2 FITC-conjugated antibodies. After FACS isolation based on the indicated gate (**blue outline**), ABCG2+ sorted cells were reanalyzed to confirm enrichment (**right panels**). **Numbers**, percent of total evaluated cells.
doi:10.1371/journal.pone.0015183.g002

unsorted IGR37 cells (0.13 grams vs. 1.4 grams, respectively; $p<0.01$) 2 months after the injection of the cells (Table 1). ABCG2- sorted IGR39 cells did not exhibit any macroscopic masses at this timepoint (Table 2), while ABCG2+ IGR39 cells produced tumors with a 3.6-fold increased mass compared to unsorted IGR39 cells (Table 2; $p<0.01$).

Single cell suspensions were obtained from all tumor xenografts and analyzed after one to two passage by means of flow cytometry to detect ABCG2-expressing cells. Consistent with previous observations with WM115 human melanoma cells and melanoma biopsies [4], ABCG2 expression was undetectable in all the tumor xenograft cell preparations (data not shown).

### CXCR6 and melanoma formation

IGR37 and IGR39 cell populations exhibited 10.6% and 9.6% frequencies of expression for CXCR6, respectively (Fig. 4). CXCR6+ and CXCR6- subpopulations of IGR37 and IGR39 cells were sorted and injected into NOD-SCID mice (n = 3 mice per cohort), as was done for cells sorted on the basis of ABCG2 expression. Surprisingly, as shown in Fig. 5 and Table 2, after only 21 days from the injection of CXCR6+ IGR37 cells, significant tumors appeared that were on average 1.8-fold greater in mass than tumors that arose from unsorted IGR37 cells ($p<0.01$). Moreover, no tumors were detected from CXCR6- injected cells (Table 2 and Figure 6) after even 2 months. Furthermore, the CXCR6- cells did not show any mass even after 4 months. Similar results were obtained for IGR39 cells. CXCR6+ cells produced tumors with 2.5-fold greater mass compared to unsorted cells; and no mass was detected for the CXCR6- cells (Table 2). As for ABCG2, CXCR6+ cells were also undetectable in tumor xenografts by flow cytometry (data not shown).

Finally, we analyzed the percentage of CXCR6+/ABCG2+ double-positive cells for both cell lines. As shown in Fig. 7, 6.7% and 3.1% of IGR37 and IGR39 cells, respectively, were double-

Table 2. Mean tumor weight ± SD (n = 3).

| Cell line | Mean±SD | Time after injection |
|---|---|---|
| IGR39 Unsorted | 0.13±0.01 | 60 days |
| IGR39 ABCG2+ | 0.47±0.01* | 60 days |
| IGR39 ABCG2- | no mass | 60 days |
| IGR39 CXCR6+ | 0.33±0.01* | 21 days |
| IGR39 CXCR6- | no mass | 21 days |
| IGR39 AB/CX | 0.53±0.01 | 21 days |
| IGR37 Unsorted | 1.4±0.3 | 60 days |
| IGR37 ABCG2+ | 2.5±0.3* | 60 days |
| IGR37 ABCG2- | 1.2±0.2 | 60 days |
| IGR37 CXCR6+ | 2.5±0.2* | 21 days |
| IGR37 CXCR6- | no mass | 21 days |

doi:10.1371/journal.pone.0015183.t002

positive for ABCG2 and CXCR6. When we injected ABCG2+/CXCR6+ double positive IGR39 cells into NOD-SCID mice, tumors arose that had on average 4-fold greater mass compared to tumors developed from unsorted cells (p<0.01; Table 2). Like IGR39 cells sorted based on CXCR6 expression alone, tumors from IGR39 ABCG2+/CXCR6+ cells required only 21 days for tumor detection (Table 2).

### Melanoma phenotype of tumor xenografts

The multiparametric immunohistochemical analysis of ABCG2+ IGR37 cells xenografts (Table 3 and Figure 8) revealed that the tumor originating from these cells displayed an homogenous expression of the differentiation antigens HBM45, HMW-MAA, and of the endothelin B receptor, and an heterogenous expression of the Melan-A antigen. No specific staining for the ET-1 peptide could be documented. All cell associated progression antigens evaluated were homogenously expressed. The extracellular matrix macromolecule tenascin was detected in the tumor graft as an extensive interstitial network entrapping cell nests of variable size. Among the foetal antigens analyzed, no detectable levels of NY-ESO were found while a weak expression of MAGE antigens was observed. Similar results have been obtained for ABCG2+IGR39 cells (data not shown).

### The CXCR6+/ABCG2+ phenotype in human melanoma biopsies

Although the evaluated melanoma cell lines were derived from human tumor tissues, the newly described CXCR6+/ABCG2+ cell phenotype might have been limited to cultured tumor cells. Therefore, we investigated 4 uncultured human melanoma positive lymph node biopsy specimens for the presence of cells with the CXCR6+/ABCG2+ phenotype. All the four biopsy specimens contained a small subpopulation of cells ABCG2+ (12,2%, 8,4%, 2,65%, >0,1%). One of these three also contained a small population CXCR6+ (0.25%) and a small population double positive ABCG2+/CXCR6+ (0,3%). Fig. 9 shows the results of the flow cytometry analysis of the CXCR6+/ABCG2+ positive specimens.

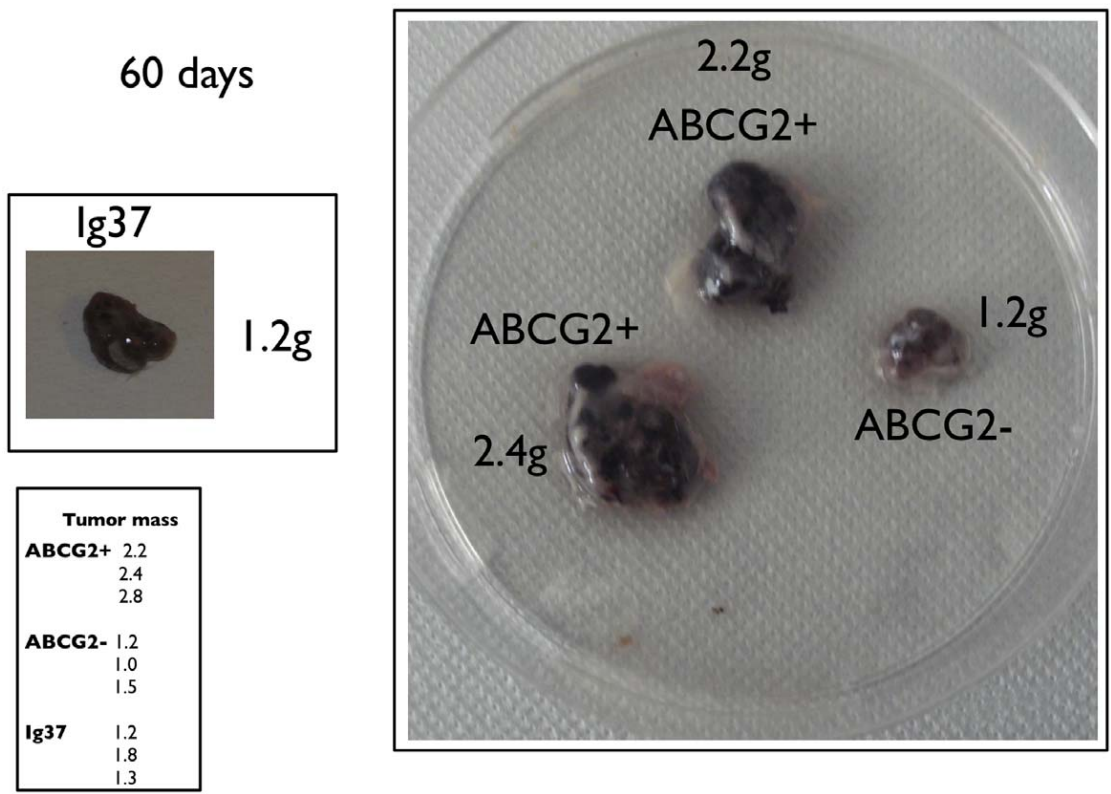

Figure 3. Tumor xenografts derived from ABCG2+ and ABCG2- IGR39 cells. $5\times10^5$ unsorted, ABCG2+ or ABCG2- sorted cells were injected subcutaneously in five-week-old NOD-SCID mice (3 mice for each experimental condition). After 60 days, the tumor mass was excised weighed and photographed.
doi:10.1371/journal.pone.0015183.g003

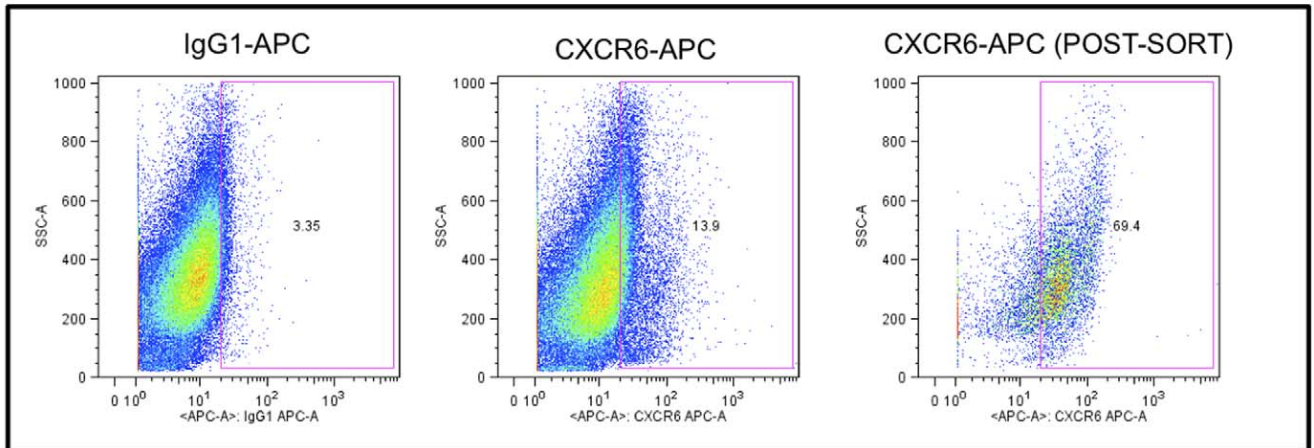

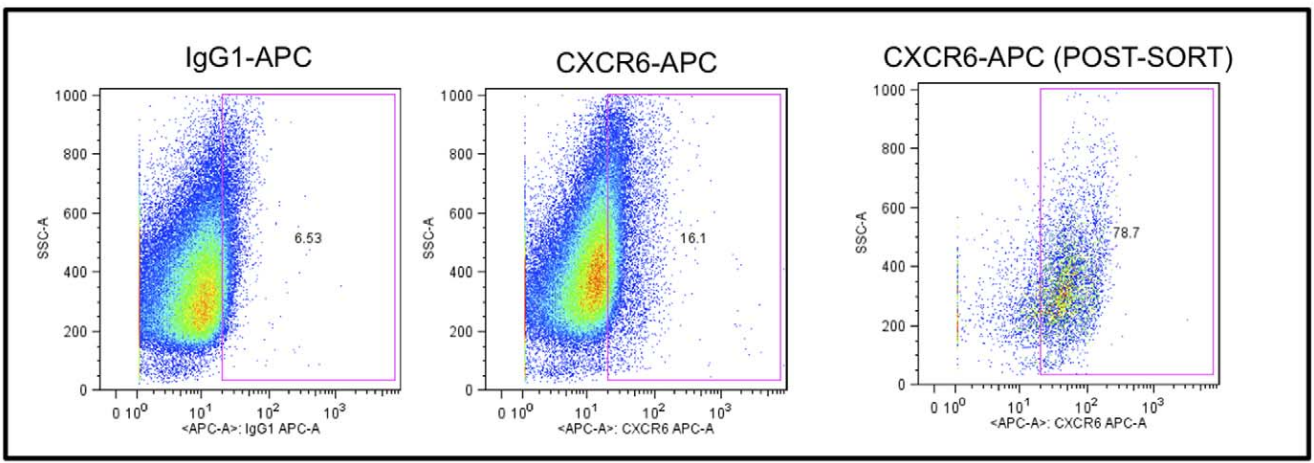

**Figure 4. Flow cytometry detection and sorting of CXCR6+ subpopulations from cultures of human melanoma cell lines.** Single marker CXCR6 analyses. Primary melanoma IGR39 cells and metastatic melanoma IGR37 cells were incubated with the indicated fluorescent APC-conjugated antibodies and analyzed by flow cytometry as described in *Materials and Methods*. **Y-axes**, side scatter; **x-axes**, relative fluorescence intensity. **Left panels**, analyses with isotype (IgG) control antibodies; **middle panels**, analyses with anti-CXCR6 APC-conjugated antibodies. After FACS isolation based on the indicated gate (**blue outline**), CXCR6+ sorted cells were reanalyzed to confirm enrichment (**right panels**).
doi:10.1371/journal.pone.0015183.g004

### CXCR6 function and CXCL16 levels in melanoma cell lines

In a recent paper, binding of CXCR6 by the cleaved soluble fragment of its membrane-bound ligand CXCL16 ("sCXCL16") was shown to enhance the proliferation of carcinoma cell lines [11]. We investigated the effects of the free sCXCL16 ligand on growth of IGR37 and IGR39 cells.

The levels of CXCL16 in the medium by unsorted cells of both cell lines and in corresponding CXCR6+ cells were assayed using ELISA (Fig. 10). The level of CXCL16 detected was extremely low (0.06 OD units ±0.03, n = 8). No significant difference was observed between the unsorted cell and CXCR6+ sorted cells after 3 days of culture (Fig. 8).

According to the described methods, both cell lines were incubated with sCXCL16 for 4 or 7 day. At the end of the incubation, the cells were trypsinized and counted. As shown in Fig. 11, by day 7, both cell lines exhibited a significant increase ($p<0.01$) that averaged 1.4-fold in cell number in response to sCXCL16 addition.

### Melanoma associated antigen (MAA) expression and ABCG2 versus CXCR6 expression

The expression profile of MAA belonging to the Cancer Testis Antigens (CTA) family (MAGE-A1, -A2, -A3, -A6, GAGE 1–2, GAGE 1-6, SSX-2, SSX 1–5, and NY-ESO-1) and of HMW-MAA was performed by RT-PCR. As shown in Table 4, similar detection patterns for ABCG2+ and ABCG2- subpopulations were observed for both cell lines. In addition, the treatment with the DNA hypomethylating agent 5-aza-2'-deoxycytidine was able to induce a *de novo* expression of MAGE-A3, MAGE-A4, GAGE 1-2, GAGE 1-6 and SSX 1-5 in both ABCG2$^+$ and ABCG2$^-$ IGR37 melanoma cells (Table 5).

With respect to CXCR6+ and CXCR6- cell subpopulations in IGR37 and IGR39 cell populations, as well as the double-positive

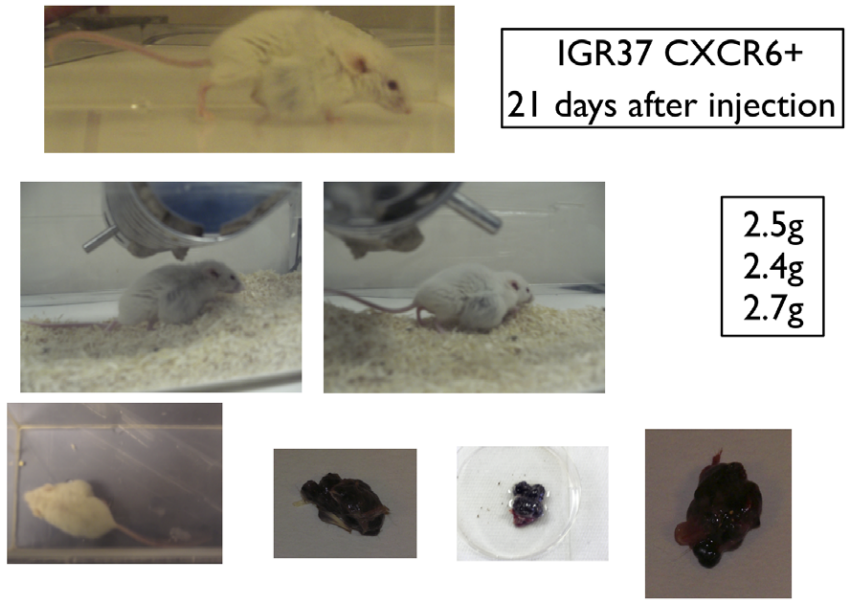

**Figure 5. Tumor xenografts derived from CXCR6+ IGR37 cells.** $5 \times 10^5$ CXCR6+ sorted cells were injected subcutaneously in five-week-old NOD-SCID mice (3 mice for each experimental condition). After 21 days, the tumor mass was excised weighed and photographed.
doi:10.1371/journal.pone.0015183.g005

ABCG2+/CXCR6+ IGR37 cells, there was an overlap of the CTA family and of HMW-MAA (Table 6). Furthermore, in the CXCR6+ IGR37 tumor xenograft, all the MAA family proteins overlapped with the exception of CXCR6+ IGR37 cells for tyrosinase, Mart-1, and Gp100 (Table 6).

## Discussion

A fundamental problem in cancer research is identifying the cell type that is capable of sustaining neoplastic growth. There is evidence that the majority of cancer cells are clones and that cancer cells represent the progeny of a single cell (reviewed in [6]). However it is unclear which cells possess the tumor-initiating cell function, maintaining the tumor growth. The idea of cancer stem cell theory is that specific cells (*i.e.*, cancer stem cells, CSCs) possess the required regenerative properties that lead to tumor formation [6]. Another important point, up to now unclear, is the origin of the putative CSCs [6]. An important way to demonstrate that a specific subpopulation can sustain the tumor growth is to use xenograft models for human cells or syngenic models for murine cells. However, an important criticism about, in particular, the xenograft model is that it is an artificial model, being, for instance, in that heterologous environmental factors are not considered [6]. On the other hand, another crucial point is to identify the markers that are able to define a cancer/initiating stem cell subpopulation.

In the present paper, we have identified a new biomarker which is associated with asymmetric self-renewal, the chemokine co-receptor CXCR6. In addition, to the role of CXCR6 in the immune response [10,12], more recently, CXCR6 expression has been reported in some human tumors, including melanoma [12]. Comparing two parental human melanoma cell lines, IGR39 (primary) and IGR37 (metastatic), the latter was more aggressive for tumor xenograft formation compared to primary melanoma IGR39 cells. This difference in tumorigenic potential was reflected in the CXCR6-sorted subpopulations. Actually, CXCR6+ cells resulted in larger tumor mass in a signficantly shorter time period, as compared to unsorted cells. Furthermore, CXCR6- cells did not result in any significant, observable tumor mass. Since CXCR6 expression was found to be linked to asymmetric self-renewal that is typical of TSSCs, MCSCs seem to be derivative of tissue-specific melanocyte stem cells. Although, it is postulated that asymmetric self-renewal by TSSCs must be ablated for them to become tumor precursors [9,15], such mutated cells may still maintain markers associated with their previous capacity for asymmetric self-renewal, even after this TSSCs function is altered by mutations.

Several ABC transporters have been found to be over-expressed in cancer cell lines cultured under selective pressure [6]. In particular, in regards to melanoma, ABCB5 is overexpressed in melanoma and ABCG2 is expressed by a sub-fraction of CD133-positive melanoma cell population [4,5]. Herein, we show, for the first time, a relationship between ABCG2 and asymmetric self-renewal. In fact, ABCG2+ cells gave a bigger tumor mass than ABCG2- cells in immunodeficient mice. However, CXCR6-expressing cells were more aggressive for tumor formation than ABCG2-expressing cells, suggesting that the subpopulation CXCR6+/ABCG2+ cells may be the most potent MCSCs.

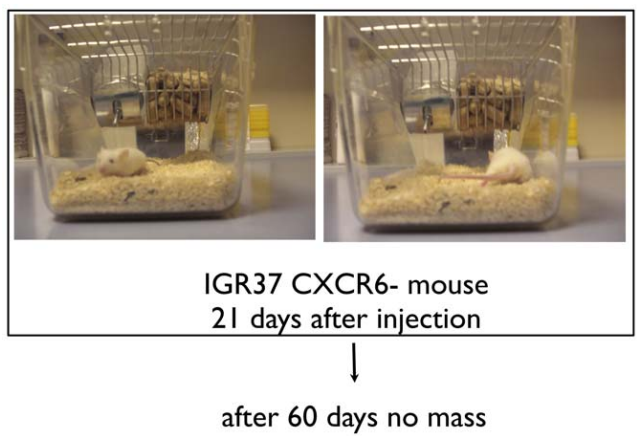

**Figure 6. Tumor xenografts derived from and CXCR6- IGR37 cells.** $5 \times 10^5$ CXCR6- sorted cells were injected subcutaneously in five-week-old NOD-SCID mice (3 mice for each experimental condition). CXCR6- cells did not yield tumors.
doi:10.1371/journal.pone.0015183.g006

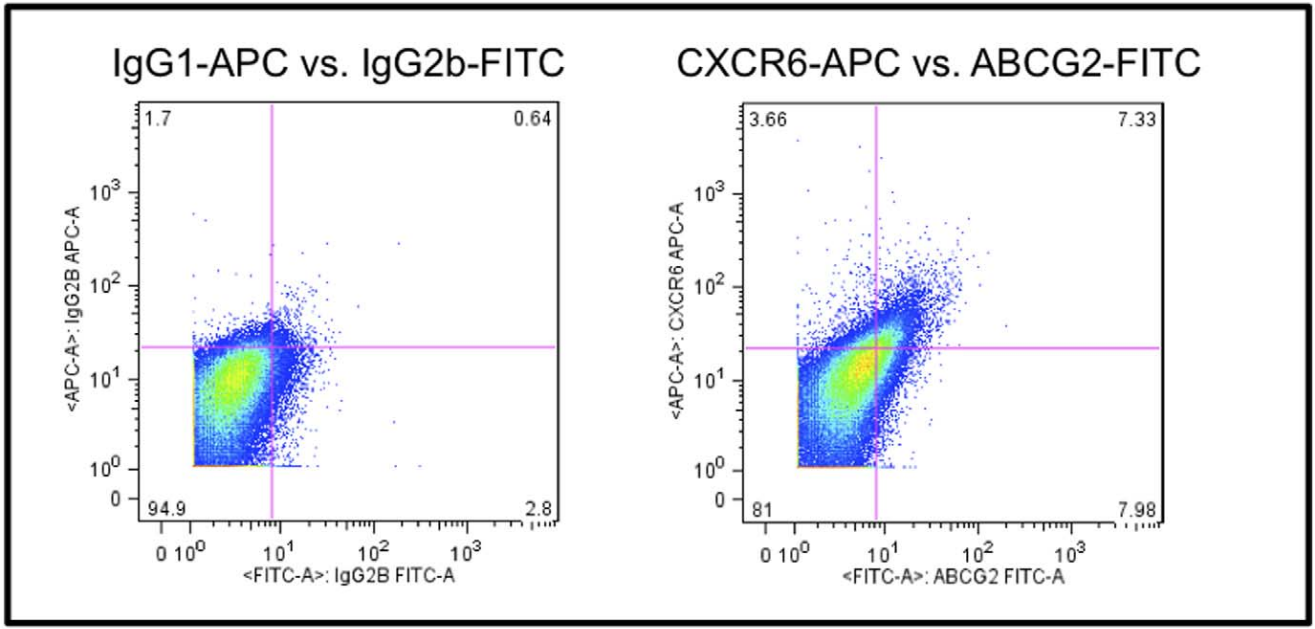

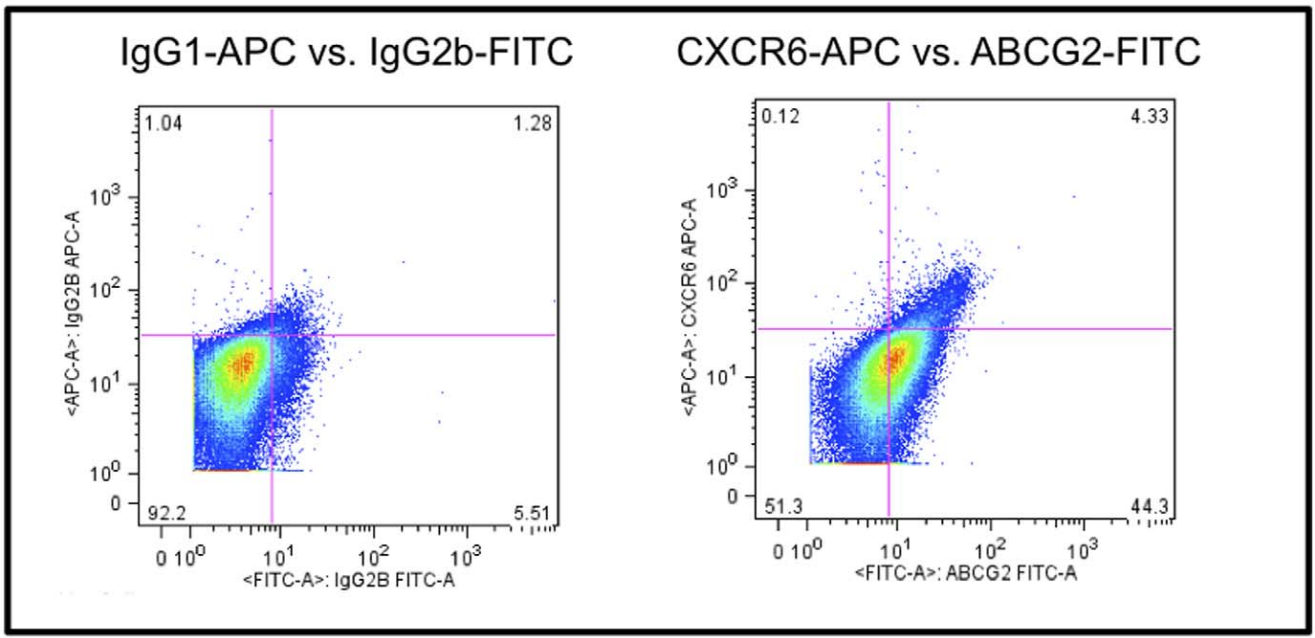

Figure 7. Flow cytometry detection and sorting of CXCR6+ subpopulations from cultures of human melanoma cell lines. Dual marker analyses for CXCR6 and ABCG2. Primary melanoma IGR39 cells and metastatic melanoma IGR37 cells were incubated with the indicated fluorescent APC-conjugated antibodies and analyzed by flow cytometry as described in Materials and Methods. **Left panels**, bivariate fluorescence intensity analyses with respective APC- and FITC-conjugated isotype (IgG) control antibodies; **middle panels**, bivariate fluorescence intensity analyses with respective APC- and FITC-conjugated CXCR6-specific and ABCG2-specific antibodies. **Numbers**, percent of total evaluated cells.
doi:10.1371/journal.pone.0015183.g007

On the other hand, an interesting question is why neither ABCG2+ nor CXCR6+ cells were detected in tumor xenografts. They might be too rare, or they might not receive the required environmental input for expression in immunodeficient mice. In

**Table 3.** Antigenic Phenotype of ABCG+ IGR37 xenograft as detected immunohistochemistry.

| Differentiation | Antigens | Progression antigens | Cancer testis antigens |
|---|---|---|---|
| HMW-MAA | positive | CD9 positive | NY-ESO-1 negative |
| HBM-45 | positive | c-met positive | MAGE-A positive |
| Melan-A | positive | HER-3 positive | |
| ET-Br | positive | Tanascin positive | |
| ET-1 | negative | | |

doi:10.1371/journal.pone.0015183.t003

fact, if we cultured tumors derived from ABCG2+ or CXCR6+ cells their expression returned in culture after one to two passages [4] (data not shown). In this regard, it is important to emphasize that our analysis of human melanoma biopsies confirmed the presence of a ABCG2+/CXCR6+ subfraction in human tissues. This finding is significant for ruling out the possibility that the ABCG2+/CXCR6+ cell phenotype exists only in cell culture. Thus, its seems more likely that expression of the proteins is prevented by some unknown mechanism in immunodeficient mice, or other species in general.

Targeting of MCSCs may present a novel and more effective clinical approach for human melanoma treatment. However, their specific antigen profile may hamper the clinical outcome of immunotherapeutic strategies for these patients. In this context, we investigated the extent MCSCs profile for specific tumor-associated antigens (TAAs) currently utilized as therapeutic targets in the clinical setting. Actually, the lack of the appropriate therapeutic targets on MCSCs might in fact result in the long-term resistance to treatment, sustained by the outgrowth of TAA-negative melanoma cell clones. In this context, the emergence of TAA-negative cells able to evade immune control has been reported in melanoma patients undergoing immunological treatment [26]. Cancer testis antigens (CTAs) encompassing large families of methylation-regulated and strictly related TAAs, first identified in human melanomas [27], represent the most promising therapeutic targets currently utilized in melanoma patients. Therefore, herein, we have addressed the following question: Can TAAs, currently utilized as targets in melanoma immunotherapeutic protocols, be used to target ABCG2-positive or CXCR6-positive cells? For both ABCG2 and CXCR6, an overlap in the expression profile of the investigated TAA and of HMW-MAA was observed between positive and negative subpopulation derived from the same parental cell cultures. In addition, the treatment with the DNA hypomethylating agent 5-aza-2'-deoxycytidine was able to induce a de novo expression of MAGE-A3, MAGE-A4, GAGE 1-2, GAGE 1-6 and SSX 1-5 in both ABCG2+ and ABCG2- melanoma cells derived from the same parental cell culture. Therefore, these data suggest the potential clinical effectiveness of TAA-based immunotheraputic strategies, alone or combined with DNA hypomethylating drugs, for more effective eradication of MCSC and non-MCSC neoplastic cells in patients with cutaneous melanoma. These findings show that CXCR6 identifies a more discrete subpopulation of cultured human melanoma cells with a more aggressive melanoma cancer stem cell (MCSC) phenotype than cells selected on the basis of ABCG2 expression. The fact that CXCR6 is a biomarker for TSSC asymmetric self-renewal supports the hypothesis that MCSCs originate for mutated melanocyte stem cells. Targeting of MCSCs may present a novel and more effective clinical approach for human melanoma treatment. However, their specific antigen profile may hamper the clinical outcome of immunotherapeutic strategies for these patients. In this context, we investigated the extent MCSCs profile for specific tumor-associated antigens (TAAs) currently utilized as therapeutic targets in the clinical setting. For both ABCG2 and CXCR6, an overlap in the expression profile of the investigated more promizing therapeutic targets TAA was observed between positive and negative subpopulation derived from the same parental cell cultures [26].

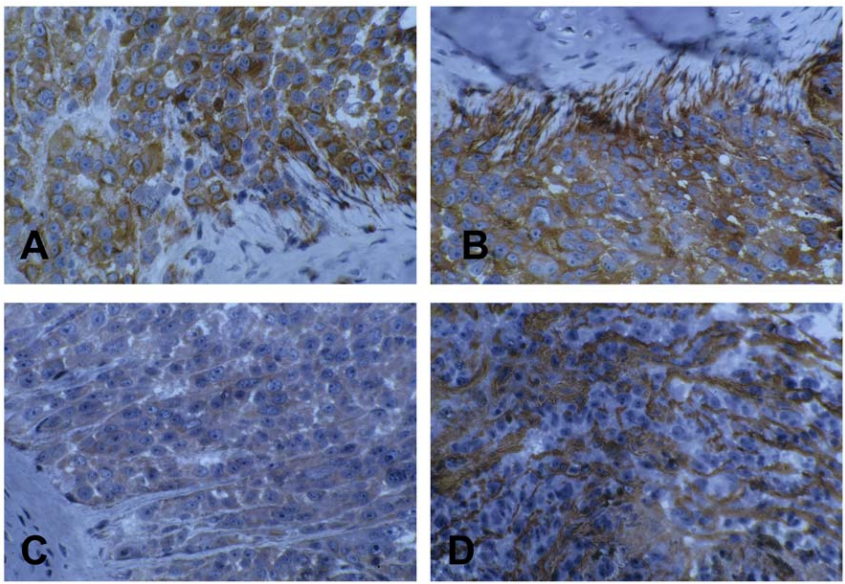

**Figure 8. Immunohistochemical analysis of ABCG+IGR37 tumor xenograft.** The tumor cell population homogenously expresses Melan-A (**A**) and the HMW-MAA (**B**) differentiation antigens. The same distribution characterizes the tumor progression antigens HER-3 (**C**) and the extracellular macromolecule tenascin (**D**). The latter characteristically delimits tumor areas of variable size. (160× magnification).
doi:10.1371/journal.pone.0015183.g008

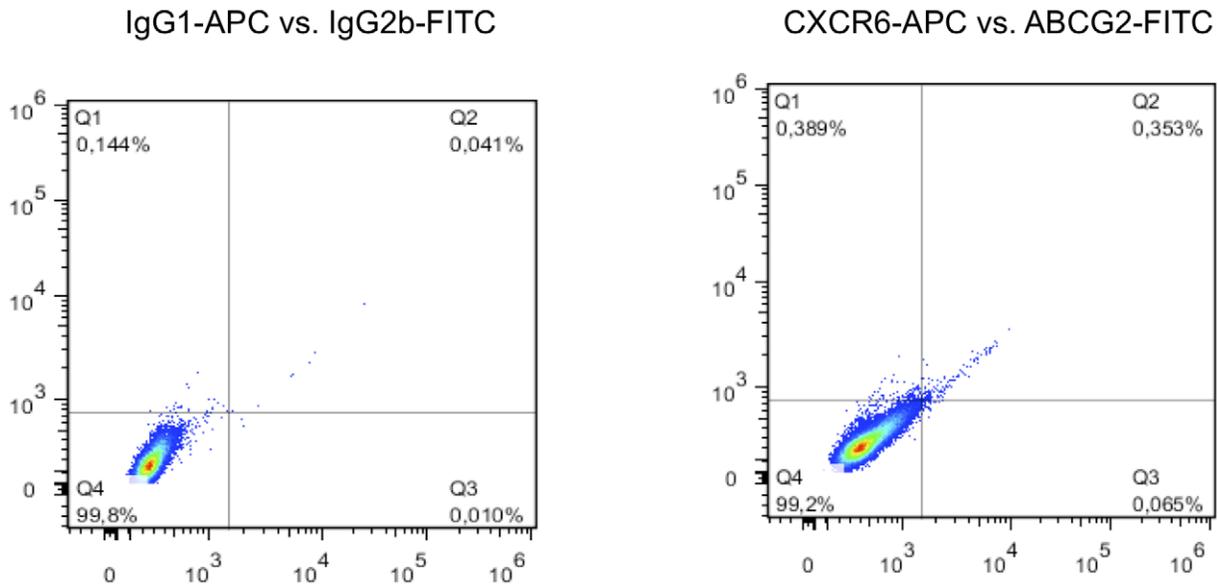

**Figure 9. Flow cytometry detection of CXCR6+/ABCG2+ subpopulations from human melanoma biopsy.** Flow cytometry analysis of a human melanoma biopsy specimen for a CXCR6+/ABCG2+ subfraction. A metastatic melanoma biopsy specimen was processed as described in *Material and Methods* and immediately analyzed by flow cytometry. **Left panel,** bivariate fluorescence intensity analyses with respective APC- and FITC-conjugated isotype (IgG) control antibodies. **Right panel**, bivariate fluorescence intensity analyses with respective APC- and FITC-conjugated CXCR6-specific and ABCG2-specific antibodies. **Numbers**, percent of total evaluated cells.
doi:10.1371/journal.pone.0015183.g009

Alltogether our findings have important implication for understanding the origin of cancer and the possible use of asymmetric self-renewal biomarkers to target more aggressive cells.

## Materials and Methods

### Ethics statement

Mice were purchased from Chrales River (Charles River Lab, Boston, MA) and housed in pathogen-free conditions at IFOM (Milan) facility. Experiments and care/welfare were in agreement with national regulations and an approved protocol by the IFOM committee approved by the National "Ministero per la Ricerca" (ID: 2/2005).

### Affymetrix oligonucleotide micro-array analysis

Whole genome expression profiles of p53-induced Ind-8 cells, p53-null Con-3 cells, and p53-induced/IMPDH II-transfected tI-3 cells [15,16,17] were compared by analyzing Affymetrix mouse

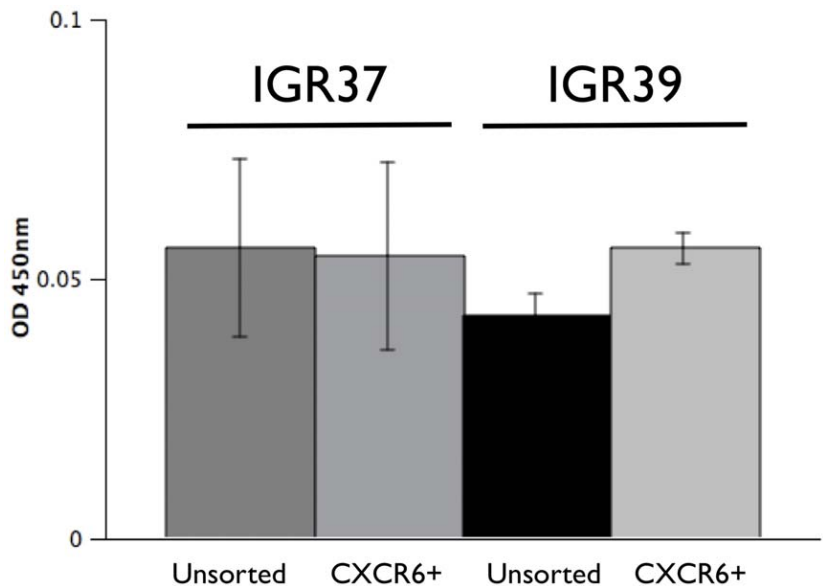

**Figure 10. CXCL16 detection in growth medium of unsorted and sorted CXCR6+ IGR37 and IGR39 cells.** 50,000 cell/well were plated in 24 multi-well plates. When the cells reached ~90% confluency (~3–4 days), medium was collected and briefly centrifuged. A 50 µl aliquot of medium was used for CXCL16 detection using the Quantikine Human CXCL16 Immunoassay (R&D Systems, Minneapolis, MN). Bars represents the mean ± S.D. of three independent experiments each carried out in quintuplicate.
doi:10.1371/journal.pone.0015183.g010

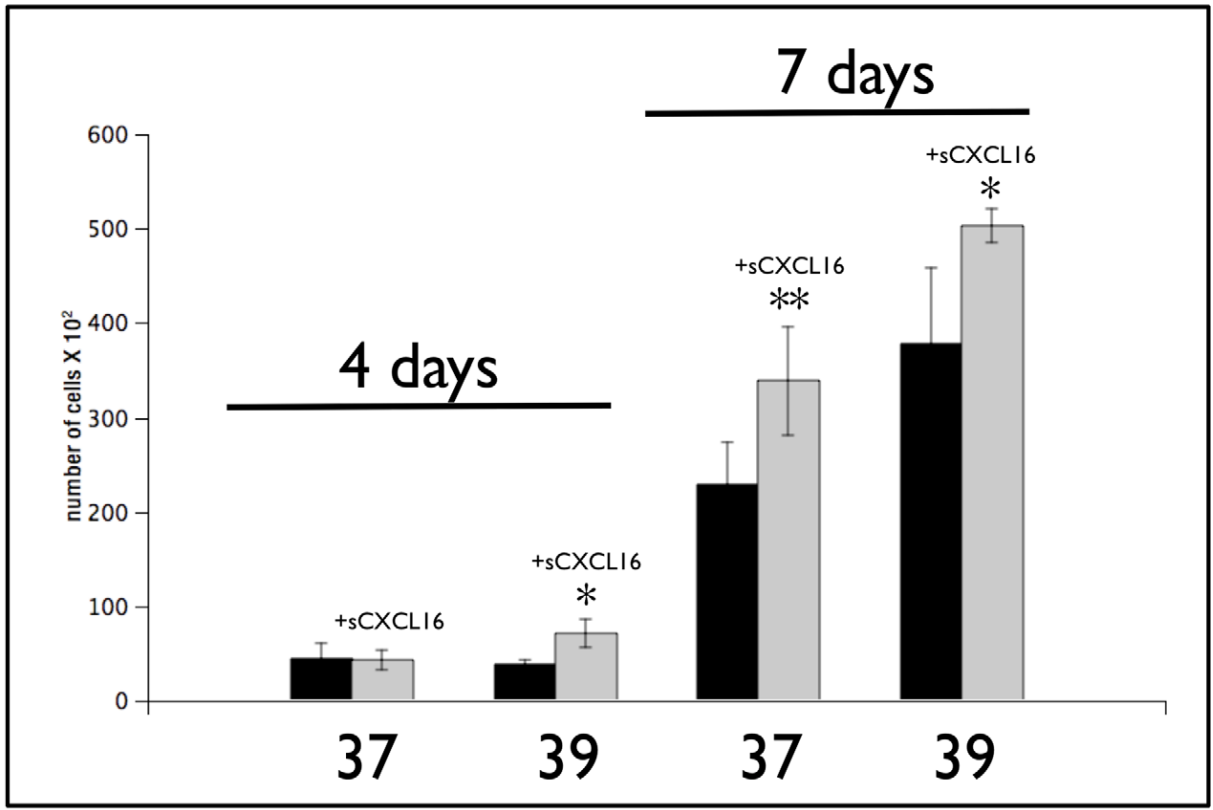

**Figure 11. Evaluation of the effect of the CXCR6 ligand CXCL16 on melanoma cell line growth.** Twenty thousand cells were seeded overnight in individual wells of 12 multi-well plates in complete medium growth condition. Recombinant sCXCL16 (100 ng/ml) was added the next day and replaced every 48 hours. On the 4th and 7th day of culture, the cells were trypsinized, and viable cell numbers determined. The bar graph represents the mean ± S.D. of three independent experiments each carried out in triplicate. *, $p<0.01$ vs. untreated cells (black bars); **, $p<0.001$ vs. untreated cells (black bars).
doi:10.1371/journal.pone.0015183.g011

**Table 4.** RT-PCR analysis of Melanoma Associated Antigens (MAA) expressed by the melanoma cell cultures IGR37 and IGR39 unsorted and sorted for ABCG2 expression.

| MAA | IGR37 | IGR37 ABCG2+ | IGR37 ABCG2- | IGR39 | IGR39 ABCG2+ | IGR39 ABCG2- |
|---|---|---|---|---|---|---|
| MAGE-A1 | - | - | - | - | - | - |
| MAGE-A2 | +/- | + | +/- | + | + | + |
| MAGE-A3 | + | + | + | - | - | - |
| MAGE-A4 | - | - | - | - | - | - |
| GAGE 1-2 | - | - | - | - | - | - |
| GAGE 1-6 | - | - | - | - | - | - |
| NY-ESO-1 | - | - | - | - | - | - |
| SSX-2 | - | - | - | - | - | - |
| SSX 1-5 | - | - | - | - | - | - |
| tyrosinase | ++ | ++ | ++ | - | - | - |
| MART-1 | ++ | + | ++ | - | - | - |
| gp100 | ++ | + | ++ | - | - | - |
| HMW-MAA | + | + | + | + | + | + |

Intensity of RT-PCR products: -, not detectable; +/-, weak; +, positive; ++, strong positive.
doi:10.1371/journal.pone.0015183.t004

**Table 5.** RT-PCR analysis of Melanoma Associated Antigens (MAA) expressed by melanoma cell line IGR39 sorted for its expression of ABCG2 and then treated with 5-AZA-CdR.

| CTA | ABCG2+ | ABCG2+ AZA 1 | ABCG2+ AZA 10 | ABCG2- | ABCG2- AZA 1 | ABCG2- AZA 10 |
|---|---|---|---|---|---|---|
| MAGE-A1 | - | - | - | - | - | - |
| MAGE-A2 | + | + | + | + | + | + |
| MAGE-A3 | - | - | +/- | - | - | +/- |
| MAGE-A4 | - | - | + | - | - | + |
| MAGE-A10 | + | + | + | + | + | + |
| GAGE 1-2 | - | - | + | - | - | + |
| GAGE 1-6 | - | - | + | - | - | + |
| NY-ESO-1 | - | - | + | - | - | + |
| SSX-2 | - | - | - | - | - | - |
| SSX 1-5 | - | - | + | - | - | + |
| PRAME | + | + | + | + | + | + |
| Tyrosinase | - | - | - | - | - | - |
| MART-1 | - | - | - | - | - | - |
| Gp100 | - | - | - | - | - | - |

Intensity of RT-PCR products: -, not detectable; +/-, weak; +, positive; ++, strong positive.
doi:10.1371/journal.pone.0015183.t005

whole genome GeneChip® 430 2.0 arrays (Affymetrix, Inc. Santa Clara, CA). Three independent cell cultures for Ind-8 and Con-3 cells and two for tI-3 cells, which were quality-controlled for respective asymmetric, symmetric, and symmetric self-renewal, were used for RNA isolation. Total RNA samples were prepared using the Trizol reagent (Invitrogen, Carlsbad, CA) and followed by a clean-up step with the Qiagen RNeasy kit (Qiagen, Valencia, CA). Total RNA quality was tested with the Agilent 2100 BioAnalyzer (Agilent Technologies, Palo Alto, CA). Five µg of total RNA was used for cDNA synthesis, and the produced cDNA was used to make biotinylated cRNA. cRNA was fragmented and hybridized onto the Affymetrix mouse whole genome GeneChip® 430 2.0 array. The arrays were washed and quantified with a fluorescence array scanner. After scanning the chips, quantification and statistics were performed using model-based expression and the perfect match (PM) minus mismatch (MM) method in the G-COS® software (version 1.0) and/or the dChip software version 2005. Data across 8 arrays were normalized by setting target intensity at 500 in the G-COS analysis. An exclusively expressed gene set for asymmetric self-renewal was selected from probe sets that were called a 'present' for all 3 independent Ind-8 cell preparations and called as 'absent' in all independent preparations of Con-3 cells (n = 3) and tI-3 cells (n = 2) by the PM/MM statistical model in the G-COS® software. The converse analysis

**Table 6.** RT-PCR analysis of Melanoma Associated Antigens (MAA) expressed by melanoma cell lines IGR37 and IGR39 sorted for its expression of CXCR6 and double positive ABCG2/CXCR6 IGR37 cells and in tumor xenograft engrafted with CXCR6+ IGR37 cells.

| CTA | IGR37CX+ | IGR37CX- | IGR37CX/AB+ | IGR39CX+ | IGR39CX- |
|---|---|---|---|---|---|
| MAGE-A1 | - | - | - | - | - |
| MAGE-A2 | + | + | + | + | + |
| MAGE-A3 | +/- | - | - | +/- | - |
| MAGE-A4 | - | - | - | - | - |
| GAGE 1-2 | - | - | - | - | - |
| GAGE 1-6 | - | - | - | - | - |
| NY-ESO-1 | - | - | - | - | - |
| SSX-2 | - | - | - | - | - |
| SSX 1-5 | - | - | - | - | - |
| PRAME | + | + | + | + | + |
| Tyrosinase | + | - | + | - | - |
| MART-1 | + | - | + | - | - |
| Gp100 | + | - | + | - | - |
| HMW-MAA | + | + | + | + | + |

Intensity of RT-PCR products: -, not detectable; +/-, weak; +, positive; ++, strong positive.
doi:10.1371/journal.pone.0015183.t006

was performed to identify a set of genes expressed exclusively during symmetric self-renewal.

All data is MIAME compliant and that the raw data has been deposited in a MIAME compliant database (E.g. ArrayExpress, GEO), as detailed on the MGED Society website http://www.mged.org/Workgroups/MIAME/miame.html.

### Evaluation of CXCR6 detection pattern in cytology assays for self-renewal pattern

**Low cell density sister pair (SP) analysis.** For low cell density SP analyses, trypsinized cells (prepared as described above) were plated at 500 cells/cm$^2$ in 2-well LAB-TEK® chamber slides in Zn-free medium (Nunc, Inc., Naperville, IL) to permit sister-sister designation based on close relative proximity. Five hours later, the culture medium were replaced with either Zn-free medium or medium supplemented to 65 M ZnCl$_2$ to induce asymmetric self-renewal. Twenty hours later, to allow for formation and cell cycle progression of sister cells, slides were washed with ice-cold phosphate buffered saline (PBS) and immediately fixed for 15 minutes at room temperature in PBS containing 3.7% formaldehyde. After three washes in PBS, the cells were permeabilized with 0.2% Triton X-100 (v/v in PBS) at room temperature for 10 minutes and afterwards washed once with PBS. Thereafter, cells were blocked with PBS containing 10% goat serum for 1 hour at room temperature. They were then incubated overnight at 4°C in a humidified chamber with an anti-mouse cyclin A monoclonal antibody (Abcam, Inc., Cambridge, UK) diluted 1:100 in PBS containing 2% goat serum. After five rinses in PBS containing 0.5% bovine serum albumin (BSA), the cells were incubated for 1 hour at room temperature with Alexa Fluor® 568-conjugated goat anti-mouse IgG (Invitrogen, Inc., Carlsbad, CA), diluted 1:300 in the blocking solution. Next, the cells were washed five times with PBS containing 0.5% BSA. For dual indirect ISIF (in situ immunofluorescence) analysis, cells stained for cyclin A detection were subsequently incubated in the same manner with anti-rat CXCR6 monoclonal antibody (R&D Systems, Inc., Minneapolis, MN) at a 1:50 dilution followed by Alexa Fluor® 488-conjugated goat anti-rat IgG (Invitrogen, Inc., Carlsbad, CA) diluted 1:300. After washing, the cells were mounted with 4'-6-diamido-2-phenylindole (DAPI)-containing VectaShield® mounting media (Vector Laboratories, Inc., Burlingame, CA). Epifluroescence images were captured with a Leica DMR microscope and Leica DC300F digital camera system. Pair-wise control analyses that omitted anti-cyclin A and anti-CXCR6 antibodies separately and together were evaluated to ensure that all detected fluorescence required these specific antibodies.

**High cell density cytochalasin D (CD) analysis.** For high density binucleated cell analysis by cytochalasin D treatment, cells were plated at 2,000 cells/cm$^2$ in 2-well LAB-TEK® chamber slides (Nunc, Inc., Naperville, IL). Thereafter cells were maintained as for SP analyses except, before fixation, they were treated with their respective media supplemented with 2 µM cytochalasin D (Sigma-Aldrich Co., St. Louis, Mo) for 14 hours to induce binucleated cell formation. The cells were then washed with ice cold PBS and immediately fixed for 15 minutes at room temperature in PBS containing 3.7% formaldehyde. Thereafter, dual indirect ISIF analyses were performed for cyclin A and CXCR6 as described above for SP analyses.

### Cell lines

Genetically engineered mouse embryonic fibroblasts (MEFs) that display TSSC asymmetric self-renewal in response to Zn-induced expression of wild-type p53 (Ind-8 cells) and congenic control vector-transfected, p53-null MEFs (Con-3 cells) were used for self-renewal pattern evaluations according to Rambhatla et al. [15]. Human IGR39 and IGR37 cells were obtained from Deutsche Sammlung von Mikroorganismen und Zellkulturen GmbH. IGR39 was derived from a primary amelanotic cutaneous tumor and IGR37 was derived from an inguinal lymph node metastasis in the same patient. The karyotype of IGR39 was tetra-pentaploid and IGR37 11% polyploidy, both showing the same rearrangements as der(1)t(1;21)(p13;q11)der(16)t(p13;q23) according to Deutsche Sammlung von Mikroorganismen und Zellkulturen GmbH. Both lines were cultured in DMEM, 15% FBS supplemented with 1% MEM vitamin, 1% MEM aminoacid, 1% antibiotics (Penicillin/Streptomycin), 1% L-glutamine (complete medium) at 37°C in a 5% CO$_2$ humidified environment.

### Tumor formation analyses and isolation of single cells from melanoma biopsy

$5 \times 10^5$ cells were injected subcutaneously into five-week-old NOD-SCID mice (Charles River Laboratories, Boston, MA). Animals were maintained on standard laboratory food ad libitum fed and free access to water.

Melanoma biopsy (four) was obtained from National Italian Institute of Tumor of Milan, Italy, after received approval from the appropriate local Institutional Review Board. Tumor mass or biopsy was excised and digested with 0.2% collagenase type I (Gibco), 0.2% bovine serum albumin (BSA; SIGMA, St. Louis, MO) for 30 minutes at 37°C on an orbital shaker.

### Flow Cytometry

Cells were analyzed for fluoroscein isothiocyanate (FITC) mouse anti-human ABCG2 (R&D Systems, Minneapolis, MN) and allophycocyanin (APC) mouse anti-human CXCR6 (R&D Systems, Minneapolis, MN) expression. All samples were analyzed using one- or two-color flow cytometry with non-specific mouse IgG used (Invitrogen, Carlsbad, CA) as isotype controls. For each flow cytometry evaluation, a minimum of $5 \times 10^5$ cells were stained and at least 50,000 events were collected and analyzed ($10^6$ cells were stained for sorting). Flow cytometry sorting and analysis was performed using a FACSAria™ flow cytometer (Becton, Dickinson and Company, BD, Mountain View, CA). Data were analyzed using FlowJo software (Tree Star, Inc., San Carlos, CA).

### SNP analysis

DNA was extracted from cells using a salting-out method [28] Genotyping for ABCG2 421 C>A (assay ID: C__15854163_70), S248P (assay ID: C__27458615_40) and F208S (assay ID: C__8826940_10) single nucleotide polymorphisms (SNPs) were performed using Validated TaqMan Genotyping Assay (Applied Biosystems). PCR reactions were carried out as follows: initial denaturation at 95°C for 10 min and 50 cycles of denaturation at 92°C for 15 seconds followed by anneal/extension at 60°C for 90 seconds. The software Opticon Monitor 2 was used in the analysis (Celbio, Italy).

### Effect of CXCL16 on cell proliferation

20,000 cells/well were seeded overnight in 12 multi-well plates in complete medium and incubated with 100 ng/ml recombinant sCXCL16 (PeproTech, Location) [12]. Live cell counts were obtained using a Vi-CELL™ Automated Cell Viability Analyzer (Beckman Coulter, Brea, CA, USA).

### Detection of CXCL16 released by cells by ELISA

50,000 cell/well were plated in 24 multi-well plates and 50 µl medium used for CXCL16 detection using the Quantikine

Human CXCL16 Immunoassay (R&D Systems, Minneapolis, MN).

### Immunohistochemistry

Tumor xenografts were snap-frozen in liquid nitrogen and kept at −70°C until processing. Four-micron cryostat sections were fixed in absolute cold acetone for 10 minutes Indirect immunoperoxidase staining employed the following murine monoclonal antibodies recognizing the indicated antigens: class I MHC antigens (clone W6/32; American Tissue Type Collection, USA), melanoma high molecular weight antigen (HMWA: clone Ep.1; Miltenji Biotec, Germany), melanosome (clone HBM-45; DakoCytomation, Denmark), Melan-A (clone A103; DakoCytomation, Denmark), CD9 (clone P1/33/2; Santa Cruz Biotech,

**Table 7.** Primers and thermal protocols utilized for the RT-PCR evaluation of CTA and differentiation antigens expression in human melanoma cells.

**RT-PCR primers**

| Gene | | Sequence | Amplicon length | Thermal cycle |
|---|---|---|---|---|
| MAGE-A1 | Sense | CGGCCGAAGGAACCTGACCCAG | 421 bp | 30 cycles of 94°C 1 min, 72°C 3 min |
| | Antisense | GCTGGAACCCTCACTGGGTTGCC | | |
| MAGE-A2 | Sense | AAGTAGGACCCGAGGCACTG | 230 bp | 30 cycles of 94°C 1 min, 67°C 2 min, 72°C 2 min |
| | Antisense | GAAGAGGAAGAAGCGGTCTG | | |
| MAGE-A3 | Sense | TGGAGGACCAGAGGCCCCC | 725 bp | 30 cycles of 94°C 1 min, 72°C 4 min |
| | Antisense | GGACGATTATCAGGAGGCCTGC | | |
| MAGE-A4 | Sense | GAGCAGACAGGCCAACCG | 446 bp | 30 cycles of 94°C 1 min, 68°C 2 min, 72°C 2 min |
| | Antisense | AAGGACTCTGCGTCAGGC | | |
| GAGE 1-2 | Sense | GACCAAGACGCTACGTAG | 201 bp | 30 cycles of 94°C 1 min, 55°C 2 min, 72° 3 min |
| | Antisense | CCATCAGGACCATCTTCA | | |
| GAGE 1-6 | Sense | GCGGCCCGAGCAGTTCA | 239 bp | 30 cycles of 94°C 1 min, 56°C 2 min, 72° 3 min |
| | Antisense | CCATCAGGACCATCTTCA | | |
| NY-ESO-1 | Sense | CACACAGGATCCATGGATGCTGCAGATGCGG | 379 bp | 35 cycles of 94°C 1 min, 59°C 1 min, 72°C 1 min |
| | Antisense | CACACAAAGCTTGGCTTAGCGCCTCTGCCCTG | | |
| SSX 2 | Sense | GTGCTCAAATACCAGAGAAGATC | 434 bp | 35 cycles of 95°C 30 s, 65°C 30 s, 72° 1 min |
| | Antisense | TTTTGGGTCCAGATCTCTCGTG | | |
| SSX 1-5 | Sense | ACGGATCCCGTGCCATGAACGGAGACGAC | 663 bp | 35 cycles of 95°C 1 min, 67°C 1 min, 72° 2 min |
| | Antisense | TTGTCGACAGCCATGCCCATGTTCGTGA | | |
| PRAME | Sense | CTGTACTCATTTCCAGAGCCAGA | 562 bp | 30 cycles of 94°C 1 min, 63°C 2 min, 72°C 3 min |
| | Antisense | TATTGAGAGGGGTTTCCAAGGGGTT | | |
| Tyrosinase | Sense | TTGGCAGATTGTCTGTAGCC | 284 bp | 36 cycles of 94°C 30 s, 60°C 1 min, 72°C 1 min |
| | Antisense | AGGCATTGTGCATGCTGCTT | | |
| MART-1 | Sense | CTGACCCTACAAGATGCCAAGAG | 602 bp | 24 cycles of 94°C 1 min, 60°C 1 min, 72°C 1 min |
| | Antisense | ATCATGCATTGCAACATTTATTGATGGAG | | |
| GP100 | Sense | TATTGAAAGTGCCGAGATCC | 361 bp | 30 cycles of 94°C 1 min, 60°C 30 s, 72°C 1 min |
| | Antisense | TGCAAGGACCACAGCCATC | | |
| HMW-MAA | Sense | CAGCAGCTCCGGGTGGTTTCAGAT | 623 bp | 30 cycles of 94°C 40 s, 56°C 40 s, 72°C 1 min |
| | Antisense | GTAGGGCCGGGCAGCATTGTAAGG | | |
| β-actin | Sense | GGCATCGTGATGGACTCCG | 615 bp | 21 cycles of 94°C 1 min, 68°C 2 min, 72°C 2 min |
| | Antisense | GCTGGAAGGTGGACAGCGA | | |

doi:10.1371/journal.pone.0015183.t007

USA), Her-3 (kindly provided by Dr A. Ullrich), c-met (clone SC-10; Santa Cruz, USA) tenascin (kindly provided by Dr. F. Malavasi), murine monoclonal antibodies to Multi-Mage-A (clone 6C1) and NY-ESO-1 (Santa Cruz, USA) cancer testis antigens (Santa Cruz, USA), endothelin B receptor (clone 39960; abcam ab., UK), and endothelin-1 (clone TR-ET.5; Affinity Bioreagents, USA). Working concentrations of the antibody were established on selected positive controls (cell lines and tumor biopsies) available in the laboratory. Staining was performed using the M.O.M. immunodetection kit (mouse primary antibodies) or Vectastain Elite ABC-Peroxidase kit (rabbit primary antibodies; Vector Lab. Burlingame, CA, USA) following the manufacturer's instructions. Nuclear counterstain employed Mayer's hematoxylin. Pictures were taken using a Leitz Orthoplan microscope.

### RT-PCR

**Cxcr6 analyses.** To validate the micro-array data, Q-RT-PCR was performed with the TaqMan® gene expression assay kit (Applied Biosystems, Foster City, CA). The kit consists of a FAM™ dye-labeled TaqMan® MGB probe and primers for detecting *Cxcr6* (Mm00472858_m1) mRNA. The expression of *Cxcr6* was normalized against the expression level of glyceraldehyde-3-phosphate dehydrogenase (*Gapdh*). One µg of total RNA from p53-induced asymmetrically self-renewing Ind-8 cells and p53-null symmetrically dividing Con-3 cells was used for reverse transcribing cDNA with the Superscript Reverse Transcriptase (RT) II kit (Invitrogen, Carlsbad, CA). The cDNA from the reverse transcription reaction was amplified by PCR to measure the FAM fluorescence of each PCR cycle using the Applied Biosystems 7700 Sequence Detection System. The reactions were performed using the manufacturer's instruction after adjusting the reaction volume as 25 µl.

**Tumor-associated antigens analyses.** Total RNA extraction and reverse transcription (RT)-PCR reactions were done as previously described [29]. The oligonucleotide primer sequences and gene-specific PCR amplification programs used are reported in Table 7. The integrity of each RNA and oligodeoxythymidylic acid-synthesized cDNA sample was confirmed by the amplification of the β-actin housekeeping gene [29]. Ten microliter of each RT-PCR sample was run on a 2% agarose gel and visualized by ethidium bromide staining. The level of expression of each gene was scored accordingly to the intensity of the specific RT-PCR product, which was obtained by densitometric analysis of ethidium bromide-stained agarose gels using a Gel Doc 2000 documentation system and the QuantityOne densitometric analysis software (Bio-Rad, Milan, Italy). Samples were scored -, no RT-PCR product detectable; +, expression level ≤10% to that of the appropriate reference cell line; and ++ expression level >10% to that of the appropriate reference cell line.

### Real-time quantitative RT-PCR analysis

Real-time quantitative RT-PCR analyses were performed as described [30]. Briefly, total RNA was digested with RNAse free DNAse (Roche Diagnostics, Milan, Italy) to remove contaminating genomic DNA. Synthesis of cDNA was performed on 1 mg total RNA using MMLV reverse transcriptase (Invitrogen, Milan, Italy) and random hexamer primers (Promega, Milan, Italy), following manufacturers' instructions. TaqMan quantitative PCR reactions were performed on 20 ng retrotranscribed total RNA in a final volume of 25 ml 1X TaqMan Universal Master Mix (Applyed Biosystems, Milan, Italy). TaqMan primers/probe sets were as follows: b-actin forward, CGAGCGCGGCTACAGCTT; b-actin reverse, CCTTAATGTCACGCACGATT; b-actin probe, FAM-ACCACCACGGCCGAGCGG-BHQ1; MAGE-A3 forward, TG-TCGTCGGAAATTGGCAGTAT, MAGE-A3 reverse, CAAA-GACCAGCTGCAAGGAACT; MAGE-A3 probe, FAM-TCTT-TCCTGTGATCTTC-MGB. Real-time measurement of fluorescent signals was performed utilizing the ABI PRISM 7000 Sequence Detection System (Applyed Biosystems) and the copy number of MAGE-A3 and of the reference gene beta-actin was established in each sample by extrapolation of a standard curve. The number of MAGE-A3 cDNA molecules in each sample was then normalized to the number of cDNA molecules of β-actin.

### Statistical analyses

Student's t-test was used to estimate the statistical confidence for differences in the weight of tumor masses that arose after the transplantation of different classes of sorted and unsorted cells.

### Acknowledgments

We thank Dr. Mario Santinami for his collaboration.

### Author Contributions

Conceived and designed the experiments: RT JS CAM LP. Performed the experiments: MN YH EC LC AC PGN MM BA LS. Analyzed the data: RT JS CAM LP. Contributed reagents/materials/analysis tools: MRN. Wrote the paper: JS CAM LP.